# Ultrafast Active Water Pump Driven by Terahertz Electromagnetic Field

Qi-Lin Zhang[1], Rong-Yao Yang[2], Chun-Lei Wang[3,4*], and Jun Hu[3,4*]

[1]*School of Mathematics-Physics and Finance and school of Materials Science and Engineering, Anhui Polytechnic University, Wuhu 241000, China*

[2]*Department of Physics, Southeast University, Nanjing 211189, China*

[3]*Zhangjiang Laboratory, Interdisplinary Research Center, Shanghai Advanced Research Institute, Chinese Academy of Sciences, Shanghai 201210, China*

[4]*Shanghai Institute of Applied Physics, Chinese Academy of Sciences, Shanghai 201800, China*

The highly efficient, easy-to-implement, long-ranged and non-destructive way to realize active pumping has been still a great challenge. Here, using molecular dynamics simulations, terahertz electromagnetic wave (TEW) is firstly employed to stimulate an active pump for water transportation by biasedly irradiated in a (6, 6) single-walled carbon nanotube (SWCNT) under no external pressure gradient. It is found that an ultrafast conductivity (up to $\approx 9.5$ $ns^{-1}$) through the pump around a characteristic frequency of 14 *THz*. The excellent pumping ability is attributed to the resonance coupling between the TEW and water molecules, in which water molecules can gain considerable energy continuously to break the binding of hydrogen bonds and the spatial symmetry. This proposed TEW-driven pump concept will offer a guide in polar molecule transport through artificial or biological nanochannels, particularly in a controllable, non-contact and large-scale process.

PACS numbers: 47.61.-k, 61.20.Ja

Keywords: molecular dynamics; water pump; carbon nanotubes; terahertz electromagnetic wave

*Introduction.* — A unidirectional continuous water transportation through nanochannels, including passive and active transport, is of great significance in many applications such as flow nanosensors [1], drug delivery [2], transmembrane transport [3], nanofiltration [4], desalination [5]. Usually, the passive water transportation through nanochannels [5-9] is realized by the hydrostatic or osmotic pressure being maintained by a large reservoir. The active transport [10] that broadly exist in biology has great realistic significance and value to develop excellent nanofluidics and regulate biological activities. However, the high efficiency, easy to implement, biocompatible and non-destructive active transport is still a significant technical challenge, partially due to the difficult efficient energy transmission through manual intervention at the nanoscale. In recent years, several novel active nanopumps have also been engineered by designing systems with vibrating carbon nanotubes [11], pre-twisting carbon nanotube (CNT) [12], charged CNT rotating [13], oscillating charges [14], or alternating hydrophobicity [15] to drive water molecules to transport continuously in single direction. All these prior studies achieved the spatial asymmetric potential of high-frequency variation by short-ranged invasive interference approaches around a single nanochannel. However, these ingenious methods are actually difficult to implement and control at the nanoscale, especially for massive porous materials and complex biochannels. Therefore, an innovative designing of active molecular pump is desired to exploit for easy bridging the practical applications.

Email: wangchunlei@zjlab.org.cn; hujun@sinap.ac.cn

Terahertz (THz) technology is nowadays of widespread applications in many fields, such as terahertz imaging, environmental monitoring, medical diagnosis, ray astronomy, broadband communication, radar etc. [16]. This is mainly attributed to the advantages of highly efficient, controllable, long-ranged and non-destructively transmitted of terahertz electromagnetic wave (TEW) [17]. Noticeably, theoretical and experimental studies show that the characteristic frequencies of water and biomacromolecule fall in the THz frequency region [18-23]. As supplementing well-established approaches, THz technology has been proven recently to be a strong tool to research protein, aqueous, and ionic solutions, which has also attracted more and more attention in the field of the artificial and biological nanochannels in recent years [24–28]. A series of studies have revealed that THz electromagnetic and mechanical vibrations can significantly affect the structure and diffusion coefficient of water inside nanochannels [29-36]. However, unfortunately, uniform irradiation of TEW will not produce pumping effect in the absence of a pressure- or density- gradient [31] in a symmetric channel. It is thus of prime significance for us to first reveal whether and how the TEW drives active transport of water molecules without external pressure gradient.

In this work, we proposed for the first time a highly efficient, controllable and long-ranged way based on the THz technology to achieve an active water nanopump by molecular dynamics (MD) simulation. It is found that this nanopump device can be stimulated with an ultrafast and continuous flow by TEW biasedly irradiating a (6,6) SWCNT. The flux maximum (even up to 9.5 $ns^{-1}$ at 14 *THz*), is comparable to the flux generated by tens of megabytes of hydrostatic pressure [9, 37], is more than three times that through the aquaporin-1 channel at a low osmotic pressure [38], and is also one order of magnitude larger than the experimentally measured flux through an SWCNT within 2 nanometers in diameter under 1 *atm.* pressure gradient [39]. Our findings are helpful for various fluid transport technologies. For example, the mechanism is expected to be used in pumping of polar molecules, transpiration in plant, steam thruster, and so on. Further, it may also be used to regulate the function of proteins being responsible for active transport in animal cells, and open up blocked artificial or biological nanochannels to promote the flow of water or blood.

*Computational methods.* — In the present setup, the overall perspective of the pumping system is displayed in Fig.1. The 50.34 *Å* long (6,6) SWCNT is perpendicularly embedded into the center holes of two parallel graphene walls with separation vacant space of 46.34 *Å*. Two water reservoirs are placed at the ends of the carbon nanotubes. Periodic simulation box of three dimensions is 3.68 × 3.83 × 9.9 $nm^3$ as the replicated unit cell containing 2091 water molecules. The center of the SWCNT with tube axis aligned along the z-direction is located at the origin of the coordinate system. All MD simulations are carried out using the large scale MD package NAMD 2.10 [40] in canonical (NVT) ensembles, constant-temperature (300 K) being maintained by Langevin thermostat with a 5 $ps^{-1}$ dumping constant. The Chemistry at HARvard Macromolecular Mechanics (CHARMM) [41] force field and TIP3P [42] water model is chosen in all the systems. The CNT-water Lennard-Jones parameters are supposed to be the constant values of $\varepsilon_{cc}$ = 0.07 $kcal/mol$, $\varepsilon_{oo}$ = 0.152 $kcal/mol$ and $\varepsilon_{HH}$ = 0.046 $kcal/mol$, where the cross interaction parameter for carbon, oxygen, and hydrogen complies with the Lorentz–Berthelot mixing criterion. The particle mesh Ewald (PME) method with a multiple time step [43] is used to calculate the electrostatic interaction. Electrostatic and Lennard-Jones interactions are computed by using a smooth (1.0-1.2 $nm$) cut-off distance. The time-step is set to be 1 $fs$, and the data were recorded every 0.5 $ps$. Duration of each simulation system is 105 $ns$: the initial 5 $ns$ was discarded for the system equilibrium, and the last 100 $ns$ samplings were used to analyze the dynamics properties. The molecular graphics program VMD was utilized as a tool for molecular visualization and resultant analysis [44]. Since the water transport properties is not sensitive to the flexibility of CNT, all carbon atoms during the simulations are treated with uncharged and frozen particles at their initial positions for simplicity in our simulations, as those in Refs [31, 45]. Following previous work, the flux is defined as the number difference of water molecules per nanosecond leaving the SWCNT from one end to the opposite end [9, 30-31].

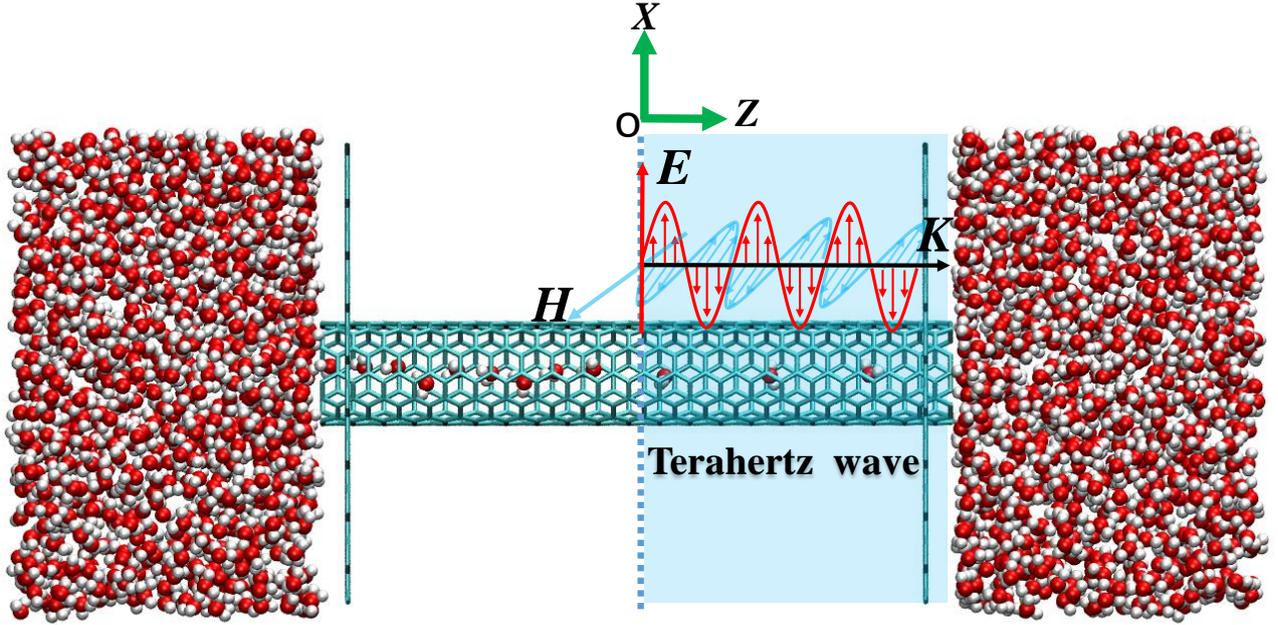

FIG. 1: (Color online) Snapshot of the simulation system. A 50.34 Å long (6,6) single-walled carbon nanotubes (SWCNT) is perpendicularly embedded into the center holes of two parallel graphene walls with separation vacant space of 46.34 Å. The distance between the bottom end of the SWCNT and the graphene is 2 Å. The balls colored by red and white represent the oxygen and hydrogen atoms of water molecules, respectively. The center of the SWCNT with tube axis aligned along the z-direction is located at the origin of the coordinate system. A uniform THz electromagnetic wave is only applied in the positive CNT space (pale blue region) along the x-direction. The red arrow, marked by $E$, stands for electric field direction.

In the present work, all simulation systems are completely symmetric potential spaces without external pressure gradient, in which water molecules can stochastically transmit in and out of CNTs because of the effect of thermal fluctuation, but no net flux occurs in a long period of time [46]. To induce a net flux, the uniform TEW is only applied in the positive CNT space with $0 \leq z \leq 2.517$ *nm* (right side region) to break the spatial symmetry of the system, as shown in Fig. 1, which can be realized in practice, e.g. focusing irradiation of TEW [47], coating nanodevices with metal [48], and wrapping the bionanochannel by the hydrogel for absorption of TEW [49]. With the THz-electric-field (TEF) direction being parallel to the *x*-direction, a periodic electric field force **F**(*t*), applied on the water molecules, is subjected by

$$\mathbf{F}(t) = q\mathbf{E}(t) = qE_0(\cos(2\pi ft), 0, 0). \tag{1}$$

where *q* represents the charge of a hydrogen or oxygen atom, *t* is the time instant, *f* stands for the frequency of TEF, and $E_0$ denotes the EF intensity. Here it should be noted that the strong TEF of order of *V/nm* generated via, e.g., difference-frequency mixing of two parametrically amplified pulse [50], can drive polar molecular rotations over random thermal motions at room temperature. In this work, unless otherwise denoted, we take the EF intensity $E_0 = 2$ *V/nm* [32]. Additionally, it should be emphasized that the magnetic force on water is so small in comparison with the electric force that it is ignored in our simulations [31].

*Results and Discussion* — MD simulations are performed for the systems under the TEF being applied in the $z \geq 0$ region (see Fig. 1) of SWCNT space at different frequencies without external pressure gradient. The flux, with regard to the TEF frequency *f*, is given in Fig. 2a, where the flux with average value of zero at *f* = 0 corresponds to the case of zero-field ($E_0 = 0$ *V/nm*). Here, the flux is calculated from the last 100 ns simulation average to generate a steady flow. Interestingly, in Fig. 2a, we can find that the direction and size of the flux are sensitive to the TEF frequency. It is observed that the flux shows different positive (+*flux*) and negative (−*flux*) values at various *f*, indicating that TEF may drive water through channels in two opposite directions dependence on the frequency. Further, within the frequency range of 6 < *f* < 24 *THz*, the +*flux* means water molecules permeate from the SWCNT left side to the right side (see the movie of 14 *THz* in Supporting Information). It is also clearly observed that the *+flux* curve shows a wide peak centered at about 14 *THz* with the maximum of about 9.5 $ns^{-1}$, which is several times larger of the conductivity than previous studies [15, 51]. On the contrary, for much lower or higher frequencies, the flux sharply falls off, and an inversion of flux direction happens at $f$ < 6 *THz* or $f$ > 24 *THz*. Fig. 2a demonstrates that a right-to-left flux (−*flux*) is generated by TEF stimulus when 1 < $f$ < 6 *THz* or 24 < $f$ < 32 *THz* (see the movie of 4 *THz* in Supporting Information). Note that there are two extreme values of ≈ 4.22 and 1.86 $ns^{-1}$ at around 4 and 26 *THz*, respectively. Additionally, once beyond the frequency band of 1 < f < 32 *THz*, there is no stable flux in Fig.2a, showing no biased transportation, similar to the zero-field case. These results show that the size and direction of water flow through nanochannels can be easily controlled through the TEF frequency. To demonstrate the robustness of the above mentioned findings, Fig. 2a (insert) shows a rough linear relationship of the cumulative number of water molecules with respect to time at zero-field, 4 or 14 *THz*, which implies that a steady-state *no-flux* (blue line), −*flux* (black line) or +*flux* (red line) is generated in a long time, respectively. As a visual reference, Fig. 2b shows the schematic diagram of water molecules through carbon nanotubes at zero-field, 4 or 14 *THz*, respectively.

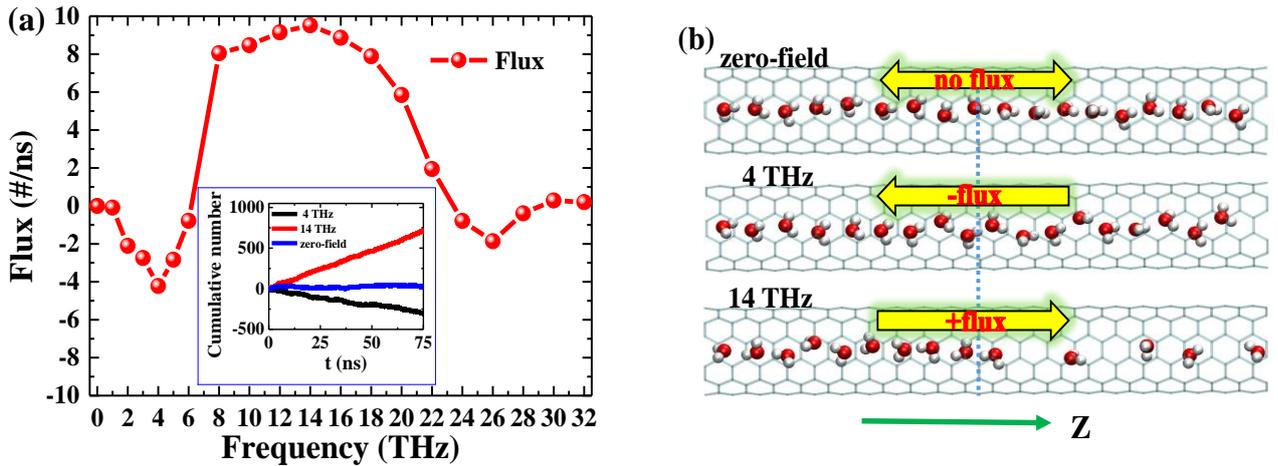

**FIG. 2: (Color online) (a) Relationship of the water flux with the electric field frequency $f$ under the intensity $E_0$ = 2 *V/nm*. The flux at $f$ = 0 corresponds to the case of zero-field ($E_0$ = 0 *V/nm*). Given in the inset is the cumulative number of water molecules with respect to time at zero-field, 4 or 14 *THz*, respectively. (b) Schematic diagram of water molecules through carbon nanotubes at zero-field, 4 or 14 *THz* respectively, where the green arrow represents the z-axis direction.**

To elucidate the detailed mechanism of water unidirectional transport under the TEF, we first investigate the average occupancy number of water molecules[14] in the left ($z$<0) and right sides ($z \geq 0$) of the SWCNT for various TEF frequency $f$, where the average is taken over the simulation time of statistics. For clarity, we define the average occupancy number of water molecules in the $z \geq 0$ (TEF region, see Fig. 3a (insert)) and $z<0$ (no TEF region) of the SWCNT as the symbols $N_L^O$ and $N_R^O$, respectively. As illustrated in Fig. 3a, the curves with the blue filled and hollow circles represent $N_L^O$ and $N_R^O$ respectively, where the zero-field result is plotted at $f = 0$. We can observe that the $N_R^O$ curve has prominent hollow profile with respect to the $N_L^O$ case in the frequency range of $6 < f < 24$ *THz*. The curve with red filled circle denotes the density ratios of $N_R^O$ to $N_L^O$, and it shows a deep valley around the interval 8-18 *THz*, where $N_R^O$ is only about half the size of $N_L^O$. As we know, water molecules can spontaneously diffuse from high density ($N_L^O$) to low density ($N_R^O$) region, and thus the left water molecules driven by the density gradient can continuously supply the right unfilled area in $6 < f < 24$ *THz*, as the visual reference of 14 *THz* in Fig. 2 b. It is a direct evidence that the density ratio exhibits an opposite tendency to that of the flux, as compared Fig. 3 a with Fig. 2 a. However, when $1 < f < 6$ *THz* and $24 < f < 32$ *THz*, although the density ratio (EQV density gradient) gradually decreases until close to 1, surprisingly we still see the opposite flux take place with two extreme values at 4 and 26 *THz*. This fact indicates that the generation and direction of flux are not uniquely determined by the density gradient.

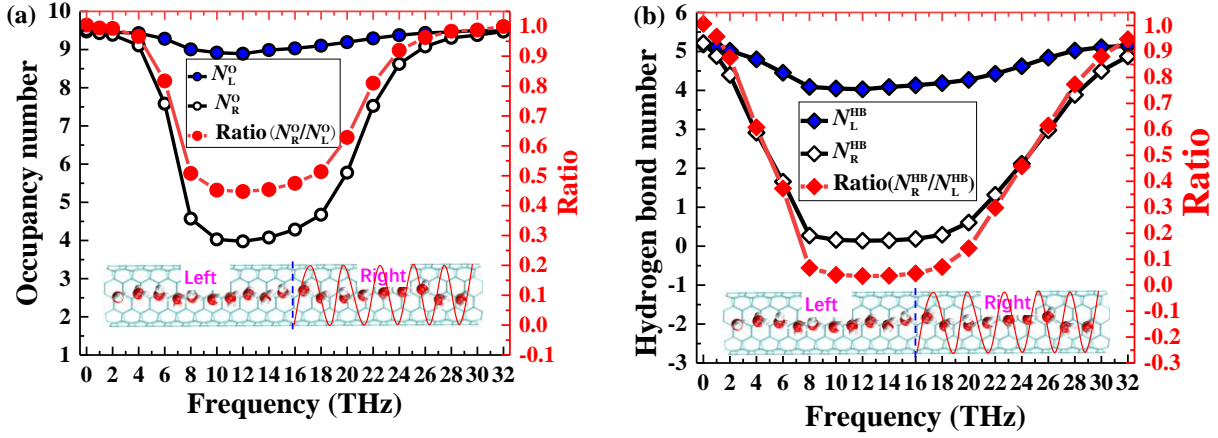

**FIG. 3:** (Color online) (a) Average occupancy number of water molecules inside the SWCNT for various electric field frequencies, where the curves with the blue filled ($N_L^O$) and hollow ($N_R^O$) circles represent the cases in the left and right sides, respectively. The curve with red filled circle denotes the density ratios of $N_R^O$ to $N_L^O$.
(b) Average hydrogen-bond (HB) number of water molecules inside the SWCNT for various electric field frequencies, where the curves with the blue filled ($N_L^{HB}$) and hollow ($N_R^{HB}$) squares represent the cases in the left and right sides, respectively. The curve with red filled square denotes the H-bond ratios of $N_R^{HB}$ to $N_L^{HB}$.

The outstanding pumping effect is closely associated with the intermolecular structures and interaction of in-tube water molecules. Next, we count the average hydrogen bond (HB) number of water molecules inside the SWCNT under the varying frequencies as plotted in Fig. 3b. The HB is defined in terms of the geometric rule of the O−O distance ≤ 3.5 Å and the bonded O −H ···O angle ≤ 30° [31, 37]. For convenience, we also define the average HB number of water molecules in the $z \geq 0$ (TEF region, see Fig. 3b (insert)) and $z < 0$ (no TEF region) of the SWCNT as $N_L^{HB}$ and $N_R^O$, respectively. Fig. 3b demonstrates that the curves with the blue filled and hollow square represent $N_L^{HB}$ and $N_R^{HB}$ respectively, where the zero-field result is plotted at $f = 0$. Similar to the occupancy number profile in Fig. 3a, in the whole frequency range $1 < f < 32$ *THz*, the $N_R^{HB}$ comparing with $N_L^{HB}$ curve has also dramatically hollow. The curve with red filled square denotes the HB ratios of $N_R^{HB}$ to $N_L^{HB}$, which shows a deep and wide valley around the frequency interval 8-18 *THz*. The dramatic drop of the HB number is associated with the breakage of HBs, due to the absorption of external energy. Our previous study and some reports have clearly revealed that the aforesaid phenomenon is ascribed to the strong resonant coupling mechanisms, characterized by librational and translational vibration modes of water molecules [30-32]. The characteristic frequency (14 *THz*) in the current study is basically consistent with our previous observations [31, 52]. Notably, the $N_R^{HB}$ is almost close to 0 in the broad frequency range of 8-18 *THz*, which indicates that water molecules can acquire enough energy in resonance to completely break the bondage of hydrogen bonds, as sketch shown in Fig. 2b. The vibration degrees of freedom and density of water molecules by the breaking of hydrogen bond are therefore markedly increased and decreased respectively, which leads to an increase in entropy and a corresponding decrease in free energy [53]. Accordingly, it is observed that the evolution of the flux is inversely proportional to the ratios of average occupancy and HB number of water molecules inside the SWCNT in the resonant frequency region of $6 < f < 24$ *THz*, with the large fluxes corresponding to small occupancy and HB numbers shown in Fig. 2a and Fig. 3. For much lower ($1 < f < 6$ *THz*) or higher frequencies ($24 < f < 32$ *THz*), although the average occupancy and HB numbers are slightly affected by TEF, the flux with a converse direction still occurs.

Up to this point, we still not fully understand why the flux direction in and near the resonance region is opposite. Now, to completely reveal biased transport details in the range of $1 < f < 32$ *THz*, we thus estimated the average interaction energies $P_W$ of water molecule at local *z* coordinate with other water molecules and all the carbon atoms in the cell system. Fig. 4a plots the $P_W$ profiles of simulation systems for zero-field (blue line), 4 (black line) and 14 *THz* (red line) respectively, where the two vertical dashed lines (pink) denote the positions of left and right ends of the SWCNT. The $P_W$ profiles are also roughly illustrated by the dashed lines of subfigures in the insert. As shown in Fig. 4a, $P_W$ of zero-field is approximately left-right symmetry except for the slight difference at the two ends of the nanotube, which is caused by the ordered alinement of water molecules in the narrow nanochannel [15]. Next, it can be observed that the 14 *THz*-$P_W$ value (red) at $z \geq 0$ (TEF region), with the barrier difference of about 12 *kcal/mol*, is significantly greater than that at $z < 0$ (no TEF region). Intuitively, water molecules tend to transport along the direction of potential energy reduction. However, in the current setup, the density gradient (see Fig. 3a) is opposite to the potential energy gradient. Arguably, the flow direction of water through the SWCNT are determined by the competition between potential energy and density gradients. Meanwhile, the water molecules from the right reservoir are difficult to enter the SWCNT from the right end with barrier of up to 12 *kcal/mol* at room temperature, resulting in that the water molecules inside the SWCNT prefer to escape from the right end. Consequently, for $6 < f < 24$ *THz*, the density gradient, over the energy gradient, plays a dominant role in driving water transportation. The overall flow thus presents a positive direction (+*flux*) from left to right (see Fig. 2a). As the density and potential energy gradient balance each other, the suppressed flux reduces to about zero at around 6 and 24 *THz* (see Fig. 2 a). For much lower or higher frequencies, the density ratios gradually decreased until close to 1, as shown in Fig. 3a. That is, for $1 < f < 6$ *THz* and $24 < f < 32$ *THz*, the biased transport direction is mainly determined by asymmetrical $P_W$. As an example, Fig. 4a displays that the 4 *THz*-$P_W$ value on the right is obviously larger than that on the left side. Water molecules prefer to migrate along the direction of decreasing potential, forming a right-to-left flux (−*flux*), as shown in Fig. 2a. Further, the competition between potential energy and density gradients leads to the occurrence of two −*flux* maxima at $f = 4$ and 26 *THz*. In addition, to investigate the atomic details of the flow asymmetry of water molecules through the SWCNT, shown in Fig. 4b are the trajectories of a water molecule in *z* direction as a function of time at 4 and 14 *THz* frequencies. The oscillating propulsion trajectories indicate that the water molecules will shuttle in some local space along the z-axis in a short time, but for a long time, the overall movement direction is still unidirectional. The reason for this phenomenon is that the evolution of the $P_W$ with the *z* coordinate is not linear for 4 and 14 *THz*, and the water molecule has to oscillate around some inflection points (e.g. $z$ = 2.5 Å at 4 *THz*) of the $P_W$.

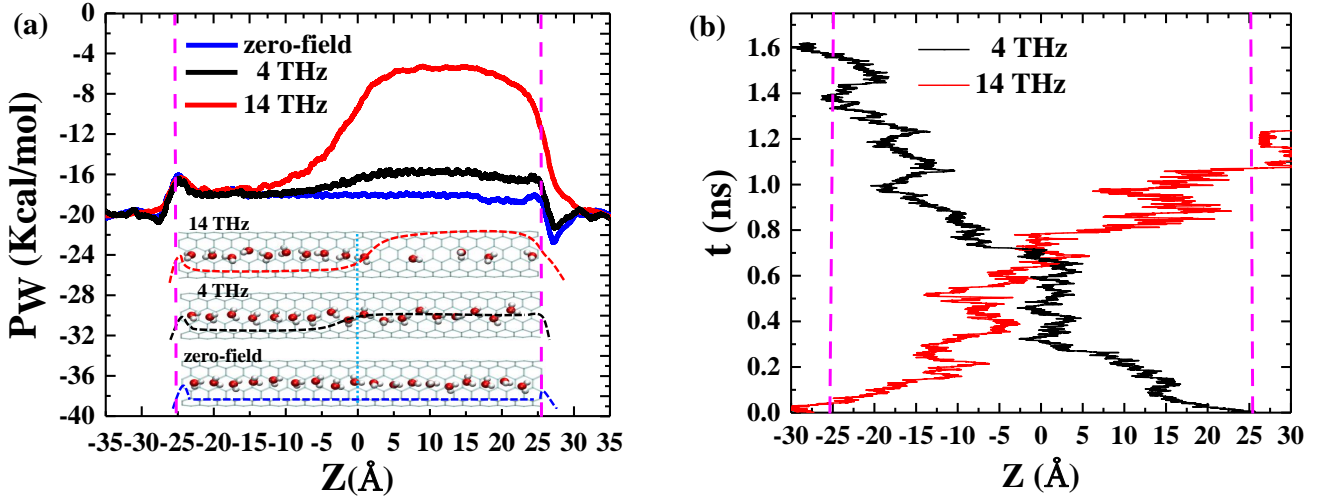

FIG. 4: (Color online) (a) Interaction energies $P_W$ of water molecule with other water molecules and carbon atoms in the simulation system as a function of z-position for zero-field, 4 and 14 *THz*. The two dashed lines (pink) stand for both ends position of the SWCNT. Insert: Schematic diagram of the suffering potential profiles (dashed line) of water molecules inside carbon nanotubes. (b) The trajectories of a water molecule in z direction as a function of time at 4 or 14 THz.

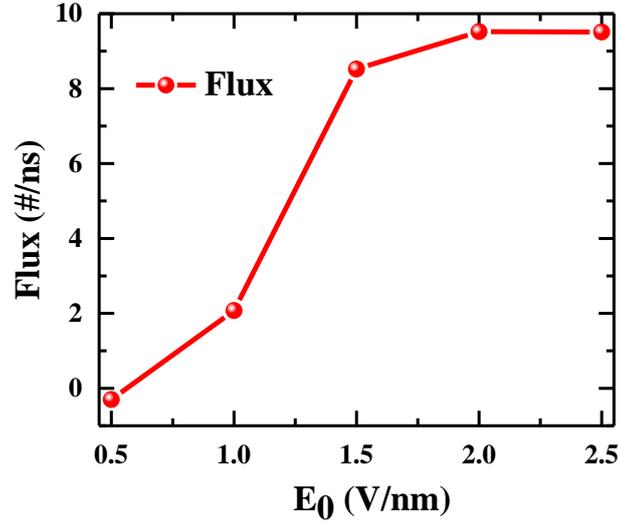

**FIG. 5: (Color online) The fluxes as a function of different electric field intensity $E_0$ at 14 *THz*.**

In the following, it is worthy to discuss the dependence of water pumping on the TEF intensity. Fig. 5 shows the net water fluxes as a function of different TEF intensity $E_0$ at 14 THz. It is found the flux increases sharply with increasing $E_0$ from 0.5 to 1.5 V/nm and almost keeps constant≈ 10 $ns^{-1}$ for $E_0$>1.5 *V/nm*, whereas the flux is almost unaffected when the electrical field $E_0$ is less than 0.5 *V/nm*. These results are attributed that water molecules inside the nanochannel are still unavoidably affected by thermal motion at room temperature, but the small intensity TEF cannot provide enough torque to suppress the thermal fluctuations of water molecules [14, 15]. Additionally, for $E_0$>1.5 *V/nm*, water molecules can completely obey the resonant coupling of the TEF which acts like a saturated energy absorption state, so that the fluxes are not significantly different at the higher strength of TEF, for instance, at $E_0$ = 2 *V/nm* in Fig. 5.

Finally, in order to investigate the application range of THz active water pump, we have performed some additional simulations (see supporting materials for specific details). It is found that TEW can still stimulate the unidirectional flow of water in large diameter nanochannel, cyclic peptide nanotubes [54], and confined two-dimensional space. Meanwhile, we also investigate effect of TEF biased irradiation at different positions on the flux. These results indicate that our proposed mechanism is universal at the nanoscale.

*Conclusions*. — To summarize, a new pumping mechanism for manipulating water biased transport through nanochannels in the absence of external pressure gradient is proposed by controlling the frequency of TEW. A series of simulations show that, when terahertz electromagnetic wave is only partially applied in a symmetrical nanochannel, a steady water flow can be excited in a single direction. This active water pump can realize ultrafast conductivity (up to $\approx 9.5\ ns^{-1}$) and controllable pumping direction by adjusting the frequency of TEW. The key physical mechanism governing the pumping is explained as follows: due to the resonances, arising from librational and translational modes of in-tube water molecules coupling with the TEW, the average interaction energies, occupancy and hydrogen-bond numbers have changed significantly under the applied THz region. As a result, the system equilibrium is so markedly broken that density and potential energy gradients are simultaneously established, resulting in generating a continuous biased flow. This proposed pumping approach based on the THz technology can take place in a controllable, long-ranged, non-contact and large-scale process, which will provide an innovative idea for the development and design of active water pump in practice. These findings also open a new avenue to develop excellent nanofluidices [55] and explore possible implications for targeted delivery of molecules or ions in transmembrane channels.

*Data Availability.* — The data used to support the findings of this study are available from the corresponding author upon request.

*Acknowledgements.* — This work is supported by National Natural Science Foundation of China (12022508, 12074394 and 11604001), and overseas and domestic visiting research projects of outstanding young backbone talents in Anhui Universities (gxgnfx2020092). The computing resources is partly provided by the Supercomputer Center of University of Science and Technology of China.

*Author contributions*

Qi-Lin Zhang and Chun-Lei Wang proposed the research idea and wrote the manuscript. Qi-Lin Zhang performed the simulations and the data analysis. All authors contributed to interpreting and analyzing the results.

**Competing interests**

The authors declare no competing interests.

*Additional information.*

Supplementary information is available for this paper at xxx